# The Mythical Swing Voter

DAVID ROTHSCHILD, Microsoft Research, NYC, NY
SHARAD GOEL, Microsoft Research, NYC, NY
ANDREW GELMAN, Departments of Statistics and Political Science, Columbia University, NYC, NY
DOUG RIVERS, Department of Political Science, Stanford University, Stanford, CA

## Introduction

In the 2012 U.S. presidential election, news coverage featured months of continuous polling and analysis of swings in support for the candidates, and campaigns spent over $2.6 billion on the presidential election alone, in large part to attract support from these swing voters.[1]

But there is a puzzle: candidates appeal to swing voters in debates, campaigns target advertising toward swing voters, journalists talk about swing voters, and the polls do indeed swing—but it is hard to find people who actually have switched sides. Political scientists have debated whether such swings in the polls are a direct response to events during the campaign, or are merely predictable changes in opinion as voters become more informed about the candidates [Gelman and King 1993; Hillygus and Jackman 2003; Kaplan et al. 2012]. This debate, however, has focused on the meanings of these swings, not their existence [Kaplan et al. 2012]. Given the general belief that poll swings reflect opinion changes, the usual resolution to the puzzle is to posit that there is a pool of moderate, or otherwise persuadable voters, who switch their support from one candidate to the other as the campaign unfolds [Fiorina and Abrams 2008; Hillygus and Shields 2009].

To resolve the incongruity between this ostensibly unmovable electorate and large reported shifts in public opinion, we conducted one of the largest ever election panel studies: an opt-in poll continuously available on the Xbox gaming platform during the 45 days preceding the 2012 U.S. presidential election. In total, 750,148 interviews were conducted with 345,858 unique respondents, who indicated the candidate for whom they intended to vote each time they were interviewed. The respondents also provided demographic information about themselves prior to the first time they responded. We focus on changes in sentiment around the first presidential debate, on October 3, 2012—a period that traditional polling suggested was pivotal in the campaign—and so we restrict our analysis to the 83,283 users who first responded before that date. On average, these individuals responded on more than four of the 45 days—yielding 336,805 responses—with over 5,000 people answering at least 15 times.

## Results

We use the Xbox data to study swing voting in two steps. First, after adjusting for demographic differences between Xbox survey participants and the electorate, we confirm that the Xbox surveys track with conventional polls. Second, we further adjust our estimates based on party affiliation to demonstrate how little of the poll movement was due to actual opinion swings, and how much was an artifact of different rates of response between Democrats and Republicans.

The Xbox panel is not representative of the electorate, so to transform the raw Xbox data into meaningful estimates of voter intent, we demographically *poststratify* the raw Xbox responses to mimic a representative sample of likely voters. We use a combined model-based poststratification strategy,

---

known as multilevel regression and poststratification (MRP), that has a growing history of use in political science and statistics [Gelman and Little 1997], and in particular, is becoming a widely applied tool in the analysis of public opinion [Lax and Phillips 2009] and voting [Ghitza and Gelman 2013]. We apply MRP by first partitioning the population into thousands of categories: 2 gender x 4 race x 4 age x 4 education x 50 states plus the District of Columbia, for a total of 6,528 demographic cells.[2] We then run fitted models to yield daily estimates of candidate support for each cell, thus we overcome cell sparsity by "borrowing strength" from related demographic cells. In the final, poststratification step, daily estimates for each of the 6,528 cells are weighted by the proportion of the electorate in each cell. In this manner, the demographically over-represented individuals in our sample are effectively down-weighted in the final estimate. For transparency, simplicity, and repeatability for future elections, we assume the 2012 electorate mirrors that of the previous presidential election, and we accordingly weight the cells based on demographics collected in the 2008 exit polls. Our results are nearly identical when we repeat our analysis with the 2012 exit poll data.

Despite having been derived from a non-representative sample of Xbox users, the MRP-adjusted estimates of support are quite similar to those from standard telephone polls conducted via random digit dialing (RDD).[3] In particular, the most striking trend in these polls is the precipitous decline in Obama's support following Romney's unexpectedly strong performance in the first presidential debate, on October 3. This substantial swing in the polls was widely reported in the media to represent a real and important change in public opinion of the two candidates. This is meaningful, because advancing in the polls is widely believed to have cascading consequences for campaigns [Schlozman and Tierney 1986; Mutz 1995]. For example, the Romney campaign saw a surge in donations and volunteers in the days following the debate, ostensibly in part due to his perceived viability.[4]

But was the reported swing in support for Romney in fact real? We use stated partisanship to address this questions, as it is highly predictive of vote intention, and it is thus natural to use this information when adjusting estimates. Crucially, we assess partisanship only when respondents first take a poll, yielding fixed, initial values for each panelist, which by construction are collected prior to the first debate. Our panel design thus ameliorates the concern in traditional, cross-sectional surveys that voters who switch their support from one candidate to another may also switch their self-reported partisanship. Analogous to our earlier analysis, we apply MRP to adjust survey responses simultaneously for demographic variables and partisanship.

Figure 1 plots Obama's daily fraction of two-party support as estimated by demographics only and demographics with partisan poststratification. In contrast to estimates from demographic only poststratification (shown in gray), additionally correcting for partisanship shows that shifts in support, while not absent, are relatively modest. Whereas adjusting only for demographics yields a six-point drop in support for Obama in the first four days following the first debate, adjusting for both demographics and partisanship indicates that in actuality there was likely only a two to three point movement. More generally, throughout our polling period, adjusting for partisanship reduces swings by more than a factor of two relative to adjusting for demographics alone.[5] Notably, while controlling only for demographics suggests Romney took the lead after the first debate, additionally adjusting for partisanship shows that Obama in fact led the race throughout. Correcting the spurious swings thus qualitatively changes the narrative of the campaign, and would have ostensibly altered the allocation of resources in this multi-billion dollar race.

---

[2]This particular partitioning was effectively determined by the polling interface, where at most four multiple-choice answers were allowed per question, with the exception of state, which respondents entered as free form text.

[3]We smooth the estimates over a four-day moving window, matching the typical duration for which standard telephone polls were in the field in the 2012 election cycle.

[4]http://abcnews.go.com/blogs/politics/2012/10/obama-has-biggest-fundraising-day-ever.

[5]Our partisan-corrected estimates are effectively an upper bound on the true net movement, as controlling for more endogenous variables could further reduce apparent swings.

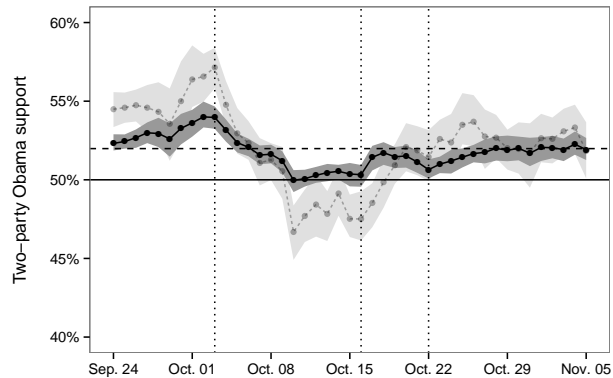

Fig. 1. Among respondents who expressed support for either Barack Obama or Mitt Romney, estimated support for Obama (with 95% confidence bands) under two different poststratification models: the dark lines plot results after adjusting for both demographics and partisanship, and the light lines adjusts only for demographics. The surveys adjusted for partisanship show less than half the variation of the surveys adjusted for demographics alone, suggesting that most of the apparent changes in support during this period were artifacts of partisan non-response.

The small shift in support that does occur is driven by people unaffiliated with major parties, moderates, and non-voters in 2008. Moreover, of the supporters Romney did gain, the majority were previously undecided; relatively few individuals switched their support from Obama to Romney.

The differences between the two lines in Figure 1 persist over time and thus cannot merely be a statistical efficiency gain produced by decreasing sampling error. Figure 2 reveals what happened: for the two weeks following the first debate, Democrats were simply much less likely than Republicans to respond to the (opt-in) Xbox poll, a fact that was not fully reflected in the demographic composition of respondents. Thus what appeared to be a major swing in support of Romney is actually be a small shift in support coupled with a large change in the relative likelihood of Democrats to respond to the poll.

What we have found is consistent with the growing literature on partisan polarization [Baldassarri and Gelman 2008; Gerber et al. 2013], but it goes against the general attitude among journalists and political scientists that poll swings represent real changes in candidate preference. In the paper we further demonstrate that this is not just an artifact of Xbox's opt-in polling design, but with careful analysis we show that partisan non-response happened in standard polling as well.

## Discussion

Looking forward, we note three implications of our findings. First, pollsters can and should poststratify on characteristics such as party identification that are associated with differential response rates. Such adjustment not only reduces the variance of estimates, but, critically, also reduces bias. While these adjustments are easiest with a panel design, it is possible to do so in cross-sectional surveys as well by making use of the assumption that the poststratifying variable (in this case, party identification) is constant (or at least varies slowly) over time [Reilly et al. 2001]. Second, starting with a non-traditional, opt-in sampling design, we have found that correcting for known demographic skew in the data yields estimates largely in line with standard, representative telephone polls, and correcting for additional covariates resulted in further improvements. The relatively low cost of our approach facilitated a survey of unprecedented scale, a fact that is relevant not just for political polling but for social research more generally. Finally, our results raise the question of whether shifts in past elections—such as the regularly occurring political convention "bounce"—are likewise attributable to partisan non-response, a question that can be studied using historical, individual-level polling data [Zaller 2002].

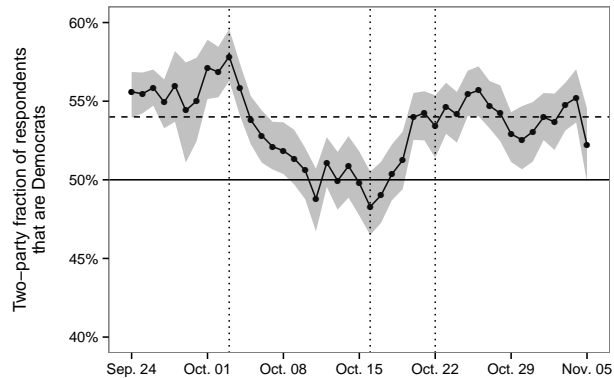

Fig. 2. Among respondents who report affiliation with one of the two major parties, the estimated proportion who identify as Democrats (with 95% confidence bands), adjusted for demographics. The dashed horizontal lines indicate the final party identification share, and the dotted vertical lines indicate the three presidential debates.